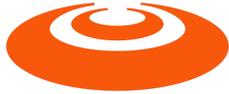

**Whitepaper**

# AI empowering research

## 10 ways how science can benefit from AI

César França
CESAR Innovation Center
Recife, Brazil

franssa@cesar.school

## Abstract

This article explores the transformative impact of artificial intelligence (AI) on scientific research. It highlights ten ways in which AI is revolutionizing the work of scientists, including powerful referencing tools, improved understanding of research problems, enhanced research question generation, optimized research design, stub data generation, data transformation, advanced data analysis, and AI-assisted reporting. While AI offers numerous benefits, challenges such as bias, privacy concerns, and the need for human-AI collaboration must be considered. The article emphasizes that AI can augment human creativity in science but not replace it.

## Introduction

This is an essay about how AI is impacting the work of scientists around the world in many different ways. However, I won't start this text explaining what AI is, or even why it turned out to be the main buzz in several technical fields, including in science, since you may already be aware of this part of the story, and science could not be that different from many other business fields. (if you don't, you can do it by quickly googling[1] or bingling[2] about)

In fact, many silent and underground changes were already in course in science when suddenly, last year, OpenAI released their now world widely famous ChatGPT[3], and the "generative AI" concept gained momentum in the mainstream media [1]. Then, it seems that everybody felt more comfortable to discuss more openly, broadly and deeply all those transformations that AI advances were causing. Recently, I read many articles showing how leading scholars, from several different fields, think about how AI could help to speed up their job [2, 3], and almost all of them evidence that AI is already in use.

---

[1] www.google.com
[2] www.bing.com
[3] https://openai.com/blog/chatgpt

When I refer to science here, I mean the whole systematic, open and collaborative processes through which people generate, share and advance what is so-called "valid knowledge". I also will sidestep the meaning of "scientific validity" here, to avoid philosophical annoyances, since I really intend to keep this text readable for as many people as possible, whatever their science education level.

However, we are proud to refer to these scientific processes as one of the most sophisticated things that has ever been created by the human kind, it is important to mention that it has never been perfect. Among the many publicly known issues that this process have (such as those related to scientometrics)[4], there is this specific behavior of scientists (and sometimes whole communities of them!) that work pretty much "click-oriented"* - i.e. they tend to focus on research subjects that are as related as possible to what the main streaming media is approaching. I am not saying that this is necessarily bad… actually, in the case for AI, we have seen an explosion of really serious and technically sound work addressing the several ways in which AI impacts science, and that has been representing maybe the most important transformation through which science has been passing, after maybe 30 years - which was when the internet, our now standard world wide network of digital communications, came about and change the way things had been done in this field since a hundred years [4].

Please, don't get me wrong. Technological and societal transformations have always been systematically around*, and they are almost always caused by advances in knowledge. Nevertheless, advances in knowledge are easier to achieve than advances in science, per se… but i have been teaching research methodology courses for university degree levels for almost 15 years by now, and I think that this time we are witnessing a truly distinct new transformation in this field. In this text, I aimed to synthesize a list of ten ways in which I believe that AI is transforming our work as scientists, followed by recent and lively examples. Surprisingly, not all of them are about speeding up our work (although the majority sure is!).

Finally, I would say that the main takeover of this article is that AI can assist human creativity in science in many different ways, but not replace it.

# 1. More Powerful Referencing

The first time I came about the concept of systematic literature review [5] was during my PhD course, in the early 2010's. That wasn't exactly a new concept, but it had recently gained attention in the field of software engineering. Even as an early stage PhD student, I could notice how transformative that thing was. What is actually special about that? It is essentially a very systematic step-by-step process of dealing with digital structure(-ish!) information (yes, an algorithm!) which enables scientists to uncover the state of art on virtually any possible existing topic.

---

[4] https://en.wikipedia.org/wiki/Scientometrics

At that moment, I was sure that it wouldn't take much time for the process to be transformed into a software tool… because it is essentially what software engineers do. As expected, it really did! [6] However, the software engineering field was just a little open door to the SLRs becoming a big thing in the field of computer science (which is where most of the AI people are in!), which could then bring much more sophistication to the process than we could have ever imagined.

Tools like these can significantly accelerate and empower our referencing work, either by tracking recent work, by finding relevant papers, by mapping knowledge areas, or even by generating whole syntheses on a given research question. AI's such as Google Scholar[5] learn what might be of interest for any scientist, and warn them about the release of relevant literature. Iris.ai[6] can help anyone to visually navigate on the literature and find papers worthy reading, as well as Elicit.org[7], which searches for similar papers to those that are provided by the user. Consensu.apps is even supposed to emulate the whole process of systematic literature reviews to quickly synthesize the answer for some specific question.

Furthermore, AI tools have been widely used to help people summarize research papers, read faster, and even interact with them in a more conversational way, as they were "talking" to the scientists (see ChatPDF for example).

## 2. Better Understanding of the Problems

During the period of my PhD research, I dealt with problems that transcended the actual borders of the software engineering field. In fact, Silvio Meira has this saying that "complex problems cannot be seen as monolithics of a knowledge area…". I was dealing with a problem that involved too much psychological knowledge, and I had to talk in person to as many psychology researchers as I could, and read as many psychology papers as possible, to develop a rich perspective about the phenomena I was questioning. I ended up eventually publishing a book on classical theories of human motivation (after the PhD was over) which is categorized in the "psychology" books at Amazon. Nevertheless, if tools like ChatGPT were on at that time, they would have saved me a significant amount of time and effort.

With the help of such conversational AI interfaces, it is possible to achieve real useful gists of complex problems, at least to overview what the common sense thinks about that. As a more practical example, in a couple of weeks ago I had to prepare a research proposal related to cybersecurity, and ChatGPT helped me on understanding common terms used in the area, showed me a general set of subtopics in which this area is currently organized, and even suggested me relevant academic and industrial problems to solve.

Of course, the ChatGPT has its own limitations as a language model, but with this information in mind, I could actually do my job more easily, running more precise searches for relevant papers on scientific databases, reading more informedly those papers, and also saving a bunch of time and effort.

---

[5] http://scholar.google.com
[6] https://iris.ai
[7] https://elicit.org

## 3. Asking better research questions

I started to work as a solo masters students supervisor at the Cesar School's Professional Master in Software Engineering in 2014. Since then, I have participated in more than a hundred bachelors, masters and phd defenses, here and in several other post-graduation programs. Believe-me, I have already lost count on how many researchers I saw starting their investigation process being completely passionate about the answers that they wanted to achieve... and I am not counting those who have just dropped out of the course after finding answers different from those that they actually wanted to achieve. Becoming a researcher demands that people are actually passionate about the questions... not about the answers.

That's why I have recently been challenging my PhD students to prompt the ChatGPT for their research questions to see what his synthetical internet-based (and neural network generated) common sense says about it. Almost always, the answer given is not the one expected by the training researcher. I must say that it is quite entertaining to see how different people react to that fact... but the most valuable part of this exercise is opening their eyes to other possible answers that their questions might get. That practice actually helps a lot to develop the ability of drawing unbiased hypotheses and to narrow down their problems into more precise questions.

## 4. Improving Research Design

"What are the most adequate research methods to answer my questions? What is the best approach to validate my results? and what issues should I care about when designing this or that specific type of inquiry?" These are some examples of prompts that researchers could use to raise insights on the moment that they are designing their research procedures.

As an example of what a student or a researcher could do, I asked ChatGPT: "I have a predictive data-based generated model. How could I scientifically validate this model?" Its answer suggested to me some very interesting approaches such as the Split Sample, Cross-Validation, Temporal Validation and the Domain Expert Evaluation. Some of these methods are explained in more technical articles that would cost me a little more time to find searching on the scientific databases. The AI even provided me with some references and guides to learn more about how to apply or use each of these methods.

Another common activity, which mainly empirical researchers deal with some frequency, is designing research instruments such as survey questionnaires, or interview and observation scripts. Generative AI can also be very helpful in doing this job too. For example, if you provide some context about your research objectives, ChatGPT-like tools could suggest some questions to assemble your script. For an even more sophisticated interaction, you could provide to the AI some guides on how to design good questions, and then check if your questions are adherent to those guidelines.

Needless to say that the AI tools, in this case, work only as a smart toolbox, that will help the scientist to discover new tools and eventually to learn how to use them. However, it is up to the scientist to make the choice of what methods will actually be used, so it is still imperative to understand why and when the best (or workable) choices are.

## 5. Generating Stub Data

Stub data refers to placeholders ,or simulated data, used during the test or prototyping of some research methods. It is used basically to mimic the behavior or characteristics of real data when the actual data is not yet available or difficult to obtain. For example, let's say that you are planning to collect a given set of variables, and design a quantitative approach to analyze your research data. You could make use of stub data to test your procedures and anticipate how the best visualizations of the results could be set up.

In the context of qualitative research, you can even pilot [8] your data collection instruments as if you were interviewing an actual subject of your research, to see if your questions work well. Furthermore, you can use the AI generated data to submit through your qualitative analysis mechanisms to see if they may work as expected.

Of course, none of these data are actually valid to achieve your results. However, they can surely be useful to validate your procedures, and that's what stub data is all about. Stub data is typically temporary and used as a temporary workaround until the actual data or functionality becomes available. Once the real data is ready, the stub data is replaced with the actual data, and the rest of the work is half way done.

## 6. Transforming Data

Qualitative research, in particular, can benefit a lot from this feature of AI tools, because this type of research takes as possible input data much richer and complex content that not always can be processed straightaway, such as pictures, documents, and either video or audio recorded interviews. Transforming those rich data into data that can be more systematically analyzed is indeed a very time and effort consuming task. Transcription has actually played a central role in spoken language analysis and representation. However, a single one hour-long audio-recorded interview, for example, can take from four hours [9] to a whole week [10] of work to become fully transcribed.

AI tools are already helping to address this problem. Image data - such as pictures or documents - can easily be described into text with MidJourney[8]. Interview recorded audio can also be transformed into text with the Whisper API[9] or the Transkriptor app[10]. Even video descriptions can be generated automatically with hypotenuse.ai.[11]

---

[8] http://midjourney.com

Again, that doesn't mean that the researcher won't need to scrutinize all the data, to make sure that the transcriptions were effectively done, and that they are quality representations of the original data, but there is no doubt that this part of the work can be made much faster. Moreover, qualitative analysis guidelines would say that to be able to achieve really meaningful and credible interpretations, researchers must develop some sort of intimacy with the data anyway [11].

For quantitative researchers, AI tools can contribute to improving the visual representation of research findings. By analyzing data patterns and trends, for example, these tools can generate interactive visualizations, infographics, or charts that effectively convey complex information. This assists researchers in presenting their results in a visually compelling and understandable manner, facilitating comprehension and reader engagement.

# 7. Analysing Data

This may be the most obvious application of AI in research. This may be not only obvious but also the most viable way for researchers to deal with large amounts of data. One of the key advantages of AI in data analysis is its ability to handle complex and large-scale datasets efficiently. AI algorithms can quickly process and analyze massive amounts of data, identifying patterns, trends, and correlations that might be challenging for humans to detect manually. By automating repetitive tasks and employing sophisticated machine learning techniques, AI systems can assist researchers in uncovering valuable insights in a fraction of the time it would take using traditional methods.

Recently, I read a research article in the area of software maintenance, in which the authors analyzed more than 10 thousand text messages recorded by programmers, using an artificial intelligence technique known as Named Entity Recognition (NER) [12]. In this technique, the researchers use a small sample of the data to train the algorithms to recognize pre-defined patterns. Then, the computer does the rest of the work. Furthermore, Atlas.TI[12], which is a very popular tool among qualitative analysts, has already released its AI-driven coding assistant that automatically performs open and descriptive coding.

The whole area of social network Sentiment Analysis[13] - a technique that is already in use in fields such as politics [13], education [14], health [15], and marketing [16] - which is only possible because of the scalable computational natural language processing algorithms [17].

---

[9] https://github.com/openai/whisper
[10] https://transkriptor.com/pt-br/
[11] https://www.hypotenuse.ai/
[12] https://atlasti.com/
[13] https://en.wikipedia.org/wiki/Sentiment_analysis

However, there are some tricks in this type of analysis too. NLP and NER may misinterpret or misclassify text data, leading to inaccurate or incorrect results. Biases present in the training data can also be learned and perpetuated by the models, leading to biased results in the analysis. These models also struggle with understanding nuances of the context, so that human expertise remains crucial in interpreting and contextualizing the results. Researchers must maintain an active role in guiding the AI systems, ensuring that the analysis aligns with their research objectives and domain-specific knowledge.

## 8. Reporting

Grammarly[14] is an online writing assistance tool that is already very popular among researchers. Its features include grammar and spelling checks, style suggestions, and even plagiarism detection. This is analogous to a copilot[15] tool for general writing. This tool became so popular that there are already several empirical investigations that agree on the positive impact of its use on the quality of general academic writing [18].

More recently, the ChatGPT has gained much attention. While ChatGPT can be useful during the research process itself, it can also assist in brainstorming, generating text to help clarify complex concepts, and even enhance the writing flow by refining the logical structure, and creating a coherent narrative. Earlier this year, the ChatGPT has even been cited as an author in some research papers, but this fact was largely criticized by the research community [19]. It culminated in the ChatGPT being banned as author, by default publication policies in many of the most respected scientific conferences [20].

Of course, this raises many ethical questions [21], and issues related to intellectual property, creative rights and authorship. Other researchers are also worried about even more complex issues, such as publication bias and researchers engaging in more dishonest practices [22]. That being said, if you plan to use creative language models such as the ChatGPT in your work, you may not only be aware of these risks, but also you would want to have a good plagiarism checker as well.

## 9. Getting feedback

Finally, before submitting your final version of the paper, why not try to predict which aspects of the text could be criticized by reviewers? As you may know, Peer review plays a critical role in validating research findings. However, not always the reviews can focus entirely on the findings, because the reviewers also have to make sure that the paper complies with superficial aspects such as format, structure, and language. Not rare, they don't even get to the main points.

Nevertheless, AI automated tools can provide initial feedback on the clarity and coherence of the manuscript, for example, and help authors to refine their texts even before submitting the paper to peer reviewing processes. Provided the existence of methodological guidelines for reporting research that uses almost all known research methods (the ACM empirical Standards[16] as a good example), conversational AI can be consulted to check if there is any lack of information in the paper.

---

[14] https://www.grammarly.com/
[15] https://github.com/features/copilot

These same practices can also be done by reviewers, so they save time to focus on the most relevant issues of the research. Also, reviewers must be equally critical with the feedback provided by AI tools, considering their own domain knowledge and research goals. A recent article [23] that investigated the potential impact of using LLMs on the peer review process concluded that LLMs can really facilitate higher quality reviews.

# 10. Uncovering new challenges that we still don't know how to tackle

As more and more AI tools keep being released at a frenetic speed [24], I stand for the authors that claim it's important to recognize that these tools serve as complements to human judgment and expertise. Human expertise, critical thinking, and domain knowledge remain essential in interpreting and contextualizing AI-driven insights. Nevertheless, the integration of AI mechanisms into the research process empowers researchers to refine their work continuously, engage in interdisciplinary collaborations, and achieve greater visibility and impact in their respective fields.

I also mentioned some challenges and limitations that we still don't know how to tackle when using AI in research, i.e. there is no consensus in academia on a single solution, or proven effective. These challenges include technical limitations, algorithmic biases, ethical concerns, intellectual property issues, regulatory compliance, and interpretability of AI-generated results. On one hand, AI algorithms are revolutionizing fields like drug discovery, genomics, astronomy, climate science, or materials science… on the other hand, these same algorithms lack auditability and explainability, which are currently fundamental requirements for rigorous scientific research [25].

Scholars still have a long path to discuss how to approach and address these challenges, to ensure the responsible and effective use of AI in their work, and I am sure that these discussions will end up in significant evolutions in this field, in the near future. As predicted in [23] AI is likely to have an even more profound impact on academia and scholarly communication in the following years.

---

[16] https://acmsigsoft.github.io/EmpiricalStandards/docs/

# About the Author

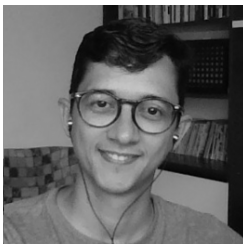

***César França*** *is PhD in Computer Science, and currently works as Head of Knowledge at CESAR, being responsible for engaging organizational efforts in the most promising investments, partnerships and research areas. At Cesar School, he is also faculty member in the Professional Master's and PhD programs in Software Engineering (http://cesar.school/), and teaches in some of the executive training programs. With extensive experience in Empirical and Experimental Software Engineering, he leads a Research Group on Social and Human Aspects in Software Engineeging (GENTE). He is also Professor of Software Engineering in the Department of Computing at the Federal Rural University of Pernambuco (UFRPE)(http://dc.ufrpe.br/) since 2015.*